\documentclass{aa}
\usepackage{graphicx,hyperref,url} 
\newcommand{\xmm}{{\em XMM-Newton}}
\newcommand{\hip}{{HIP~67522}}

\newcommand{\mearth}{$M_{\oplus}$}

\usepackage{color}

\begin{document}

\title{XUV irradiation of young planetary atmospheres. 
Results from a joint XMM-Newton and HST observation of HIP67522}

\author{A. Maggio\inst{1} 
\and I. Pillitteri\inst{1}
\and C. Argiroffi\inst{2,1}
\and D. Locci\inst{1}
\and S. Benatti\inst{1}
\and G. Micela\inst{1}}
\institute{
INAF-Osservatorio Astronomico di Palermo, Piazza del Parlamento 1, 90134 Palermo, Italy\\
\email{antonio.maggio@inaf.it}
\and
Department of Physics and Chemistry, University of Palermo, Piazza del Parlamento 1, 90134 Palermo, Italy
}

\titlerunning{XUV irradiation of HIP 67522b}
\authorrunning{A. Maggio et al.}

\abstract
{The evaporation and the chemistry of the atmospheres of warm and hot planets are strongly determined by the high-energy irradiation they receive from their parent stars. 
This is more crucial among young extra-solar systems because of the high activity of stars at early ages. 
In particular, the extreme-ultraviolet (EUV) part of the stellar spectrum drives significant processes of photo-chemical interaction, 
but it is not directly measurable because of strong interstellar absorption and a lack of sufficiently sensitive instrumentation. 
An alternative approach is to derive synthetic spectra from the analysis of far-ultraviolet (FUV) and X-ray emission lines, which allow us to estimate the missed flux in the EUV band.
}
{We performed joint and simultaneous spectroscopy of \hip\ with \xmm\ and the Hubble Space Telescope (HST) in order to reconstruct the full high-energy spectrum of this 17\,Myr-old solar-type (G0) star, 
which is the youngest transiting multiplanet system  known to date.
}
{We performed a time-resolved spectral analysis of the observations, including quiescent emission and flaring variability. 
We then derived the emission measure distribution (EMD) versus temperature of the chromospheric and coronal plasma from the high-resolution spectra obtained in X-rays with RGS and in FUV with COS.}
{We derived broad-band X-ray and EUV luminosities from the synthetic spectrum based on the EMD, which allowed us to test alternative EUV versus\ X-ray scaling laws available in the literature. We also employed the total X--EUV flux received by the inner planet of the system to estimate its instantaneous atmospheric mass-loss rate.}
{We confirm that \hip\ is a very active star with a hot corona, reaching plasma temperatures above 20\,MK even in quiescent state. Its EUV/X-ray flux ratio falls in between the predictions of the two scaling laws we tested, indicating an important spread in the stellar properties, which requires further investigation.}

\date{9 September 2024}
\maketitle 

\section{Introduction}
Since the discovery of the first extra-solar planet around a main sequence star in 1995 by \citet{Mayor1995},
the search for exoplanets has received an outstanding push forward in several directions. Huge efforts are devoted today to 
characterizing planetary atmospheres, and to understanding the formation and evolution processes leading to the observed variety of planetary masses and
sizes. Nowadays, there are more than 5000 confirmed planets and thousands of \textit{Kepler} and TESS (Transiting Exoplanet Survey Satellite) candidate planets (\citealp{Batalha2013}, \url{exoplanet.eu}, \url{doi.org/10.26134/ExoFOP5}). 
However, the available target sample is dominated by relatively old stars because of difficulties
in planet discovery around active (young) stars with current instrumentation and detection techniques.
The frequency of planets depends on their masses, sizes, and host star properties, and is a key parameter for testing planet formation and evolution models. 
On the other hand, evolutionary paths are the result of the complex interplay between physical and dynamical processes operating on different timescales, including the stellar radiation fields. 
In particular, intense high-energy irradiation from the host stars, especially at young ages, can be responsible for evaporation of exoplanet  atmospheres; this process is one of the ingredients that shapes the planet mass--radius relationship but is still poorly understood \citep{Lopez2013,Owen2013,Fulton2017,Owen2017, Fulton2018, Owen2018, Modirrousta+2020}. 
For these reasons, there are several ongoing observation programs at optical and IR wavelengths targeting relatively young stars. 
In particular, TESS is providing unprecedented opportunities for exoplanet searches around stars in young moving groups and stellar associations. 
Planets in these stellar environments are particularly important because the ages of the systems can be determined more accurately than for field stars. However, these photometric surveys can only provide measurements of planetary radii, while determination of masses still requires spectroscopic follow-up with ground-based facilities. 
For this reason, the italian GAPS collaboration is currently leading a long-term program with the HARPS-N (optical) and GIANO-B (NIR) spectrographs at the Telescopio Nazionale Galileo (TNG) in La Palma \citep{Carleo2020, Damasso2020}, and we have conducted similar Guest Observer programs with the HARPS spectrograph at the ESO-3.6m in La Silla.
The aim of these programs is to constrain planetary masses and orbital parameters of selected TESS and \textit{Kepler} young planets using the radial velocities technique. With our programs, we contributed to validating the presence of the first young TESS candidate, namely DS\,Tuc\,Ab (40\,Myr, \citealp{Benatti2019}), 
and the planetary system around the 50\,Myr K star TOI-942 \citep{Carleo2021}. 
For these and other young systems (V1298\,Tau, 10--30\,Myr, \citealp{Maggio+2022}; TOI-837, 35\,Myr, \citealt{Damasso+2024}), we  also reconstructed the photoevaporation histories.

The structure of planetary atmospheres sensitively depends on the spectral energy distribution of the stellar radiation \citep{Lammer2003}. 
While extreme-ultraviolet (EUV) photons are absorbed in the upper atmosphere, soft X-rays can heat and ionize lower layers due to the cascade of secondary electrons \citep{Cecchi-Pes2006}. 
Reliable characterization of planetary evolution requires knowledge of the whole stellar high-energy emission and its variability \citep{Sanz-Forcada2011, Locci2019}.
Planets in close orbits around young stars (t $<$ 100 Myr) are especially susceptible to irradiation effects because of the higher activity levels relative to the Sun and the stronger magnetic fields. 

Higher magnetic activity is generally accompanied by more frequent and energetic flares \citep{Davenport2016}, and higher rates of coronal mass ejection (CME) are expected as well \citep{Khodachenko2007}. 
In turn, charged particle flows linked to stellar winds and CMEs determine the size and time-dependent compression of planetary magnetospheres, and eventually may lead to stripping (erosion) of close-in planets \citep{Lammer2007}, as well as deposition of gravity waves \citep{Cohen2014}. 
A detailed characterization of the high-energy emission (1--1700\AA, hereafter XUV emission) of young stars hosting exoplanets is highly desirable with XMM-Newton or Chandra and the Hubble Space Telescope (HST) at present, in the era of the James Webb Space Telescope (JWST) and in view of the forthcoming Ariel mission in 2029.

\begin{figure}
    \centering
    \resizebox{\columnwidth}{!}{
    \includegraphics{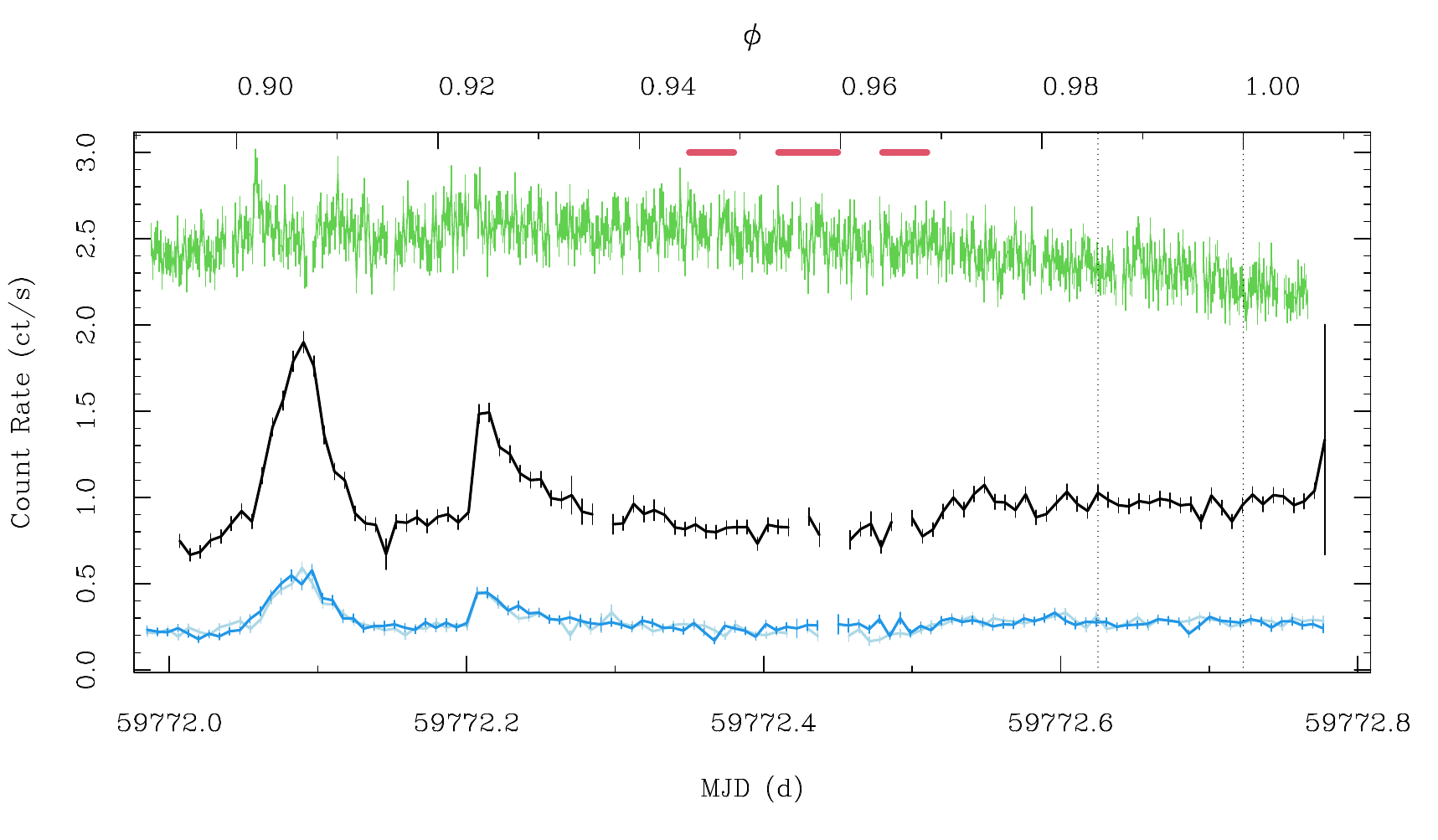}
    }
    \resizebox{\columnwidth}{!}{
    \includegraphics{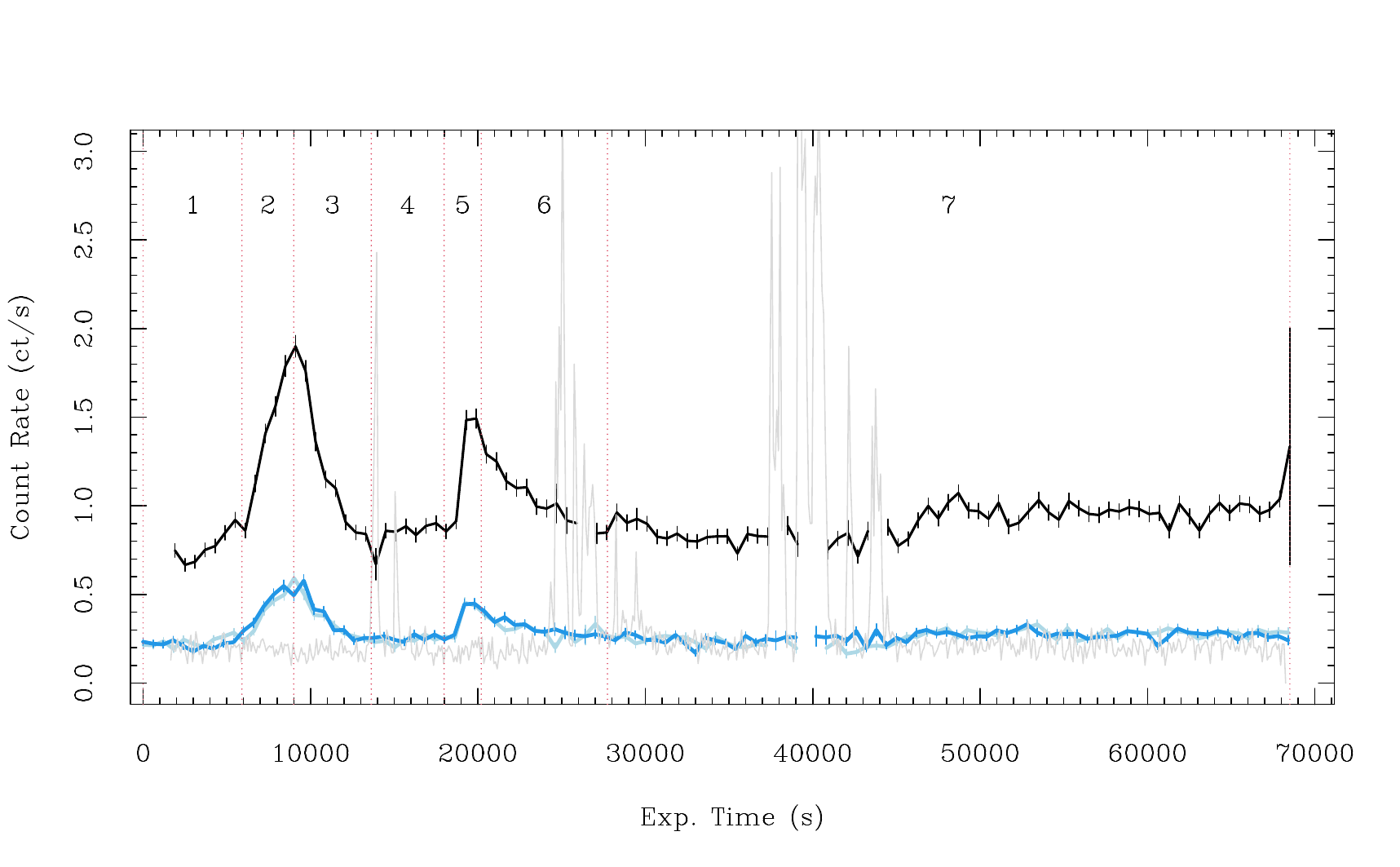}
    }
    \caption{X-ray and NUV light curves of \hip. Top: Light curves of EPIC (black pn, dark and light blue MOS 1,2) and OM (green) instruments.
    The EPIC light curves were binned at 600\,s per bin, while the OM light curve was binned at 90\,s.
    The OM rate is divided by 5 for making easier the comparison. 
    The horizontal bars on the top indicate the time of the HST/COS exposures.  
    The top axis reports the orbital phases of the planet.
    The vertical dotted lines mark ingress and mid-transit times of the planet.
    Bottom: EPIC light curves and time intervals defining flares 1 and 2 rise+peak and decay (2, 3, 5, 6) and quiescent levels (1, 4, 7). The gray light curve refers to background single events with energies between 10\,keV and\,12 keV, in the full field of view of the pn detector.}
    \label{fig:lc}
\end{figure}

HIP 67522 (HD 120411) is a member of the Sco-Cen young association (5–20 Myr), located at a distance of $124.7 \pm 0.3$\,pc. 
It hosts a Jupiter-size transiting planet discovered in the TESS survey and validated by \cite{Rizzuto2020}, with an orbital period of $\sim$6.96\,d, and a planetary radius of $10.07 \pm 0.47$ R$_\oplus$, but just a loose constraint of its mass in the range 0.18--4.6\,M$_\mathrm{J}$. Moreover, this planet is likely undergoing Kelvin-Helmholtz contraction and photoevaporation \citep{Heitzmann2021}, meaning that its mass could be lower than those of mature planets with similar radii \citep{Lopez2013}. 
By measuring the Rossiter-McLaughlin effect during transits, \citep{Heitzmann2021} also determined the orbital inclination of HIP\,67522b, and discovered that it is well-aligned in spite of its young age, thus ruling out a migration history driven by high eccentricity. The planet is also a compelling target for atmospheric characterization, and the system was indeed observed in 2023 with JWST (PI A.W.\ Mann, GO 2498).

The system also contains a second smaller planet (R$_p$ = $8.2 \pm 0.5$ R$_{\oplus}$) with a period of 14.96\,d, which was recently confirmed by \citet{Barber+2024} to be in near 2:1 mean motion resonance with HIP\,67522\,b. This discovery makes \hip\ the youngest system known to date to have two transiting planets.

In the present paper, we describe our analysis of simultaneous observations of \hip\ with \xmm\ and HST in order to acquire 
high-resolution spectra from FUV to soft X-rays, and to determine the emission measure
distribution of the optically thin plasma from the outer chromosphere to the corona, and the full XUV irradiation of the planets. 
This, in turn, is crucial input for modeling their atmospheric evaporation and photo-chemistry.

The structure of the paper is as follows: Sect.\ \ref{observations} describes the observations and the data analysis, Sect.\ \ref{results} presents the results of the analysis, and Sect.\ \ref{conclusions} contains a discussion and our final conclusions.

\section{Observations}
\label{observations}
\hip\ was observed in X-rays with \xmm\ for about 70 ks (P.I.: A. Maggio, ObsId: 0902070101) on July 11, 2022. 
Together with the European Photon Imaging Camera (EPIC), we used the Optical Monitor (OM) with the filter UVM2, whose band-pass is 200--300\,nm, in order to monitor the near-ultraviolet (NUV) emission of the star, and the Reflection Grating Spectrograph (RGS) for high-resolution spectroscopy in the band $5-35$ \AA. 

Figure\ \ref{fig:lc} shows the light curves of EPIC and OM instruments. 
Two flares are evident in the EPIC light curve, while the OM light curve shows an overall modulation on timescales of longer than the exposure time and comparable with the rotation period of the star itself ($P_{\rm rot} = 1.418 \pm 0.016$\,d, \citealt{Rizzuto2020}).
We simultaneously observed the FUV emission of \hip\ with the HST Cosmic Origins Spectrograph (COS) during three consecutive orbits  (program id: LESW01010) in order to obtain spectra with the
G130M grism in the band $1170-1420$ \AA\  (central wavelength: 1291 \AA). 
The coverage of HST orbits during the \xmm\ exposures is shown in Fig. \ref{fig:lc}.
Due to an issue with the acquisition of the target, only the second exposure was successful in providing a COS spectrum of the star. 

\subsection{Analysis of \xmm\ data}
\begin{figure}
    \centering
    \resizebox{\columnwidth}{!}{
\includegraphics[trim= 0.5cm 1.0cm 4.8cm 1cm, clip]{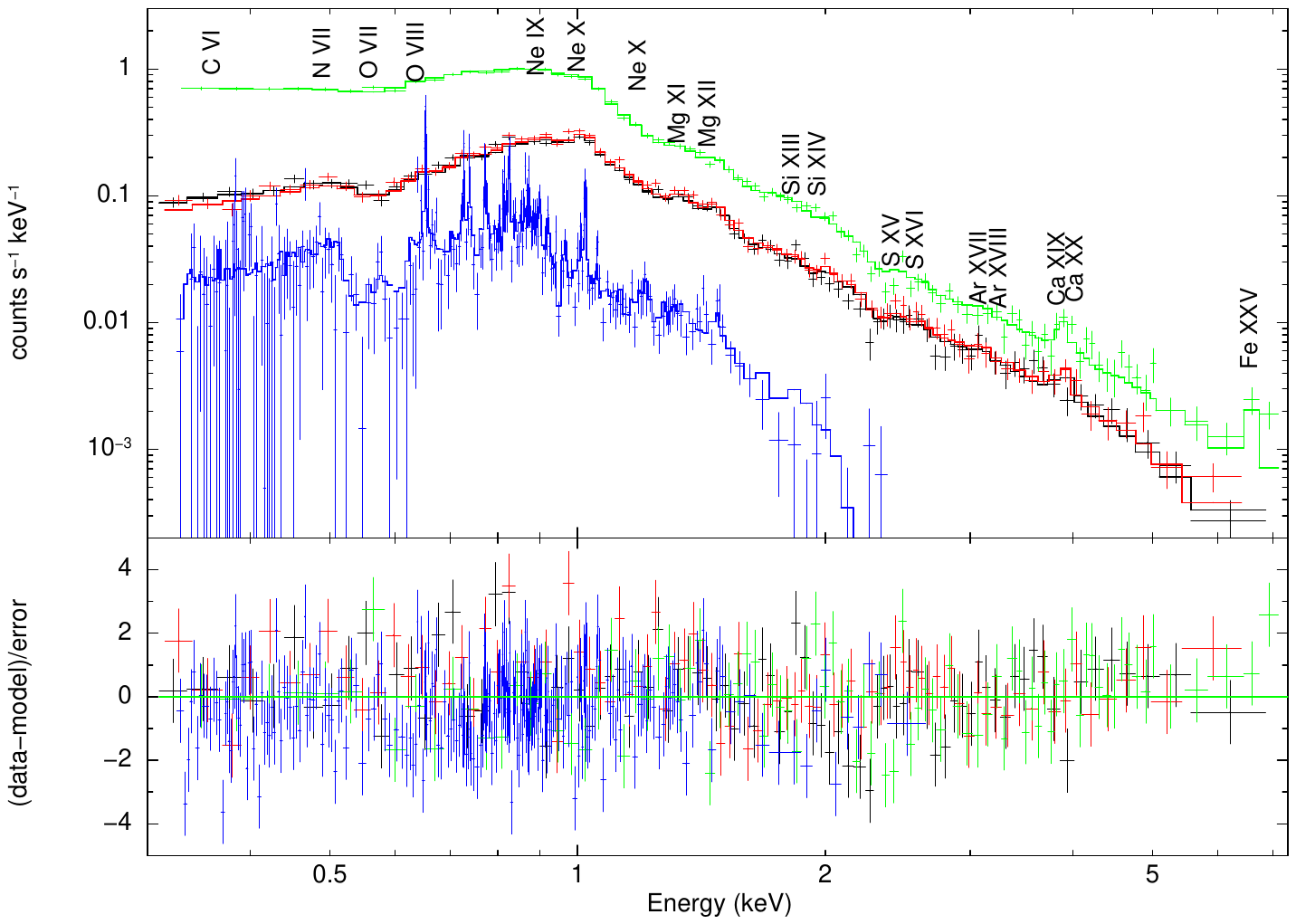}
    }
    \caption{EPIC and RGS global spectra, and best-fit 3T VAPEC model. Black and red data points for the MOS1 and MOS2, while the pn spectrum is in green, and the summed RGS1+RGS2 spectra in blue. The main complexes of H-like and He-like ions are indicated.}
    \label{fig:globalspec}
\end{figure}

The observation data files (ODFs) constituting the \xmm\ observation were retrieved from the \xmm\ archive and reduced with XMM-SAS version\ 20.0.0. 
For the MOS and pn observations, we obtained calibrated event lists in the band 0.3--8.0\,keV. The source events were extracted from a circular region with a radius of 60", while the background was extracted from a close region with similar size and devoid of other sources. From these files, we obtained source and background light curves (Fig.\ \ref{fig:lc}). Inspection of the light curve of events detected with energies $E>10$\,keV allowed us to identify and
filter out few time intervals affected by high background, as prescribed in the XMM-SAS guide. Finally, we extracted the source spectra along with the relative response files, and applied a rebinning in order to get at least 30 counts per energy bin and a signal-to-noise ratio (S/N) of at least five for each spectral bin of EPIC spectra, and 10 counts per bin in the case of RGS (Fig.\ \ref{fig:globalspec}).

\begin{figure}
    \centering
    \resizebox{\columnwidth}{!}{
\includegraphics{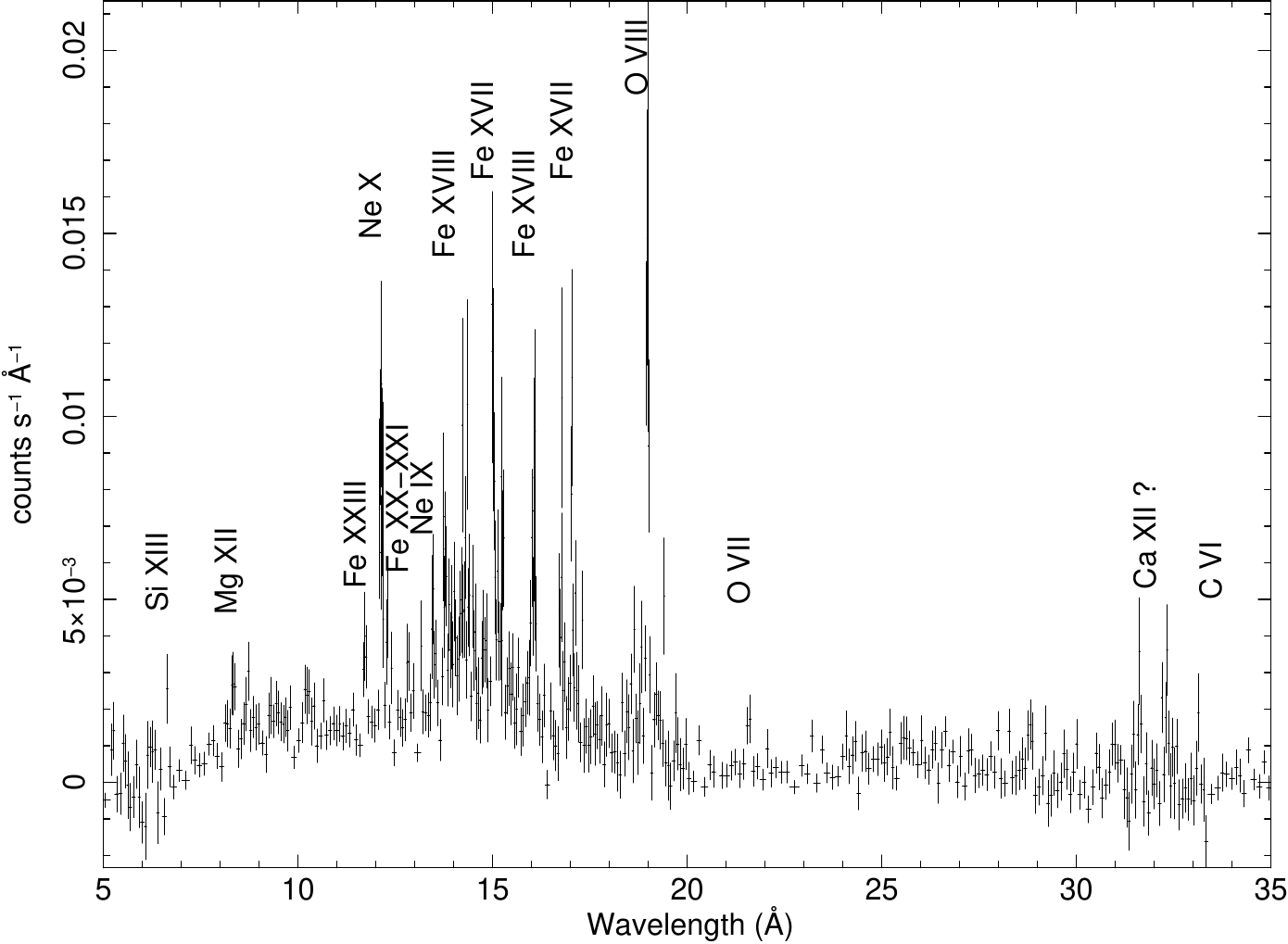}
    }
    \caption{RGS spectrum of the full exposure, with the identifications of the most prominent emission lines.
}
    \label{fig:rgs}
\end{figure}

We reduced the data from the OM and from the RGS to obtain a light curve in the band 200--300\,nm 
and a high-resolution spectrum in the $\sim 5$--35\,\AA\ wavelength interval. 
The first-order spectra of RGS1 and RGS2 were summed together in order to improve the counting 
statistics and cover the gap of the missing chips in each detector (Fig. \ref{fig:rgs}).

We then performed a global spectral analysis of the EPIC and RGS spectra jointly (Fig.\ \ref{fig:globalspec}, Sect.\ \ref{sec:globalspec}) for the entire length of the observation. The spectra were analyzed with {\sc xspec} version\ 12.12.
We adopted an optically thin plasma emission model with three isothermal components 
(VAPEC), and including absorption by the interstellar medium (TBabs model, with 
ISM abundances by \citealt{Wilms+2000}). 
The free parameters for the best-fit procedure were the interstellar H column density ($N_{\rm H}$), the temperatures (k$T$), the emission measure ($EM$) of each component, and the abundances of several elements scaled to the solar one ($Ab(Z)/Ab(Z)_\odot$, shared by all components).
The goodness of the fit was evaluated with chi-square statistics, and the uncertainties on the best-fit values were computed at the 90\% statistical confidence level with the XSPEC \textit{error} procedure.

As the X-ray emission was variable during the \xmm\ exposure (Fig.\ \ref{fig:lc}), we also performed a time-resolved spectral analysis, dividing the observation into seven intervals (Sect.\ \ref{sec:flares}). 
For each interval, we accumulated the combined spectrum of PN, MOS1, and MOS2, and built the appropriate spectral response and effective area files with the SAS tasks {\it rmfgen} and {\it arfgen}. 
High photon rates can produce pile-up and distortion of the spectrum, especially in pn. In our case, we checked 
with the SAS {\it epatplot} task that the pile-up fraction is negligible\footnote{The
 {\it epatplot} task allows us to test the presence of pile-up on each instrument,
starting from unfiltered events of MOS/pn.
For the spectrum in interval 2 (first flare peak), the nominal model of single/double
pn events in the range 0.5-2.0\,keV resulted in agreement with the observed
distribution to better than 0.5\%. Following the SAS guide, this means no
significant pile-up was recorded even during the intervals with the highest
source count rate.}, even in the interval of highest rate during the first flare, and that further corrections are not necessary.
The same conclusion applies to the global spectra introduced above.

\begin{figure}
    \centering
    \resizebox{\columnwidth}{!}{
    \includegraphics{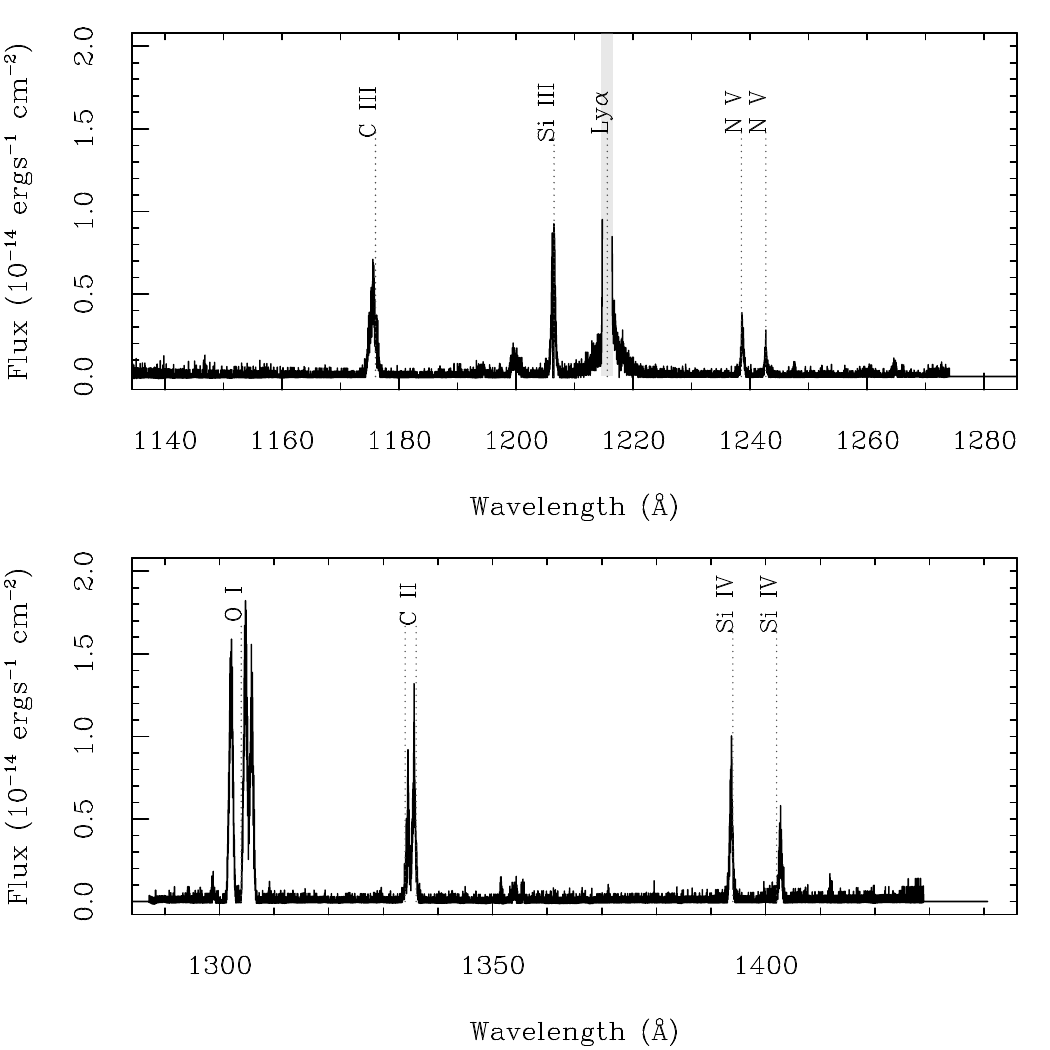}
    }
    \caption{Spectra of \hip\ obtained with COS and grism G130M (central wavelength 1291\,\AA). 
    Top: FUVB segment ($\sim 1140--1280$\,\AA). Bottom: FUVA segment ($\sim 1290--1420$\,\AA). The main lines are labeled.}
    \label{fig:spec_cos}
\end{figure}

Subsequently, the emission lines of the coronal ions in the combined RGS spectra of the global observation were measured with Pint Of Ale version\ 2.954 (PoA, \citealp{Kashyap1998,Kashyap2000}), and employed for the reconstruction of the plasma emission measure distribution versus\ temperature, EMD(T) (Sect.\ \ref{sec:EMD}), together with the chromospheric and transition region lines measured in the HST spectra.
Table \ref{tab:linefluxes} reports the measurements of the coronal line fluxes employed in the analysis.

\subsection{Analysis of HST data}
The COS spectra of the HST observations were retrieved from the HST archive, ready to be analyzed.
These spectra were acquired with COS and the G130M filter calibrated in both fluxes and wavelengths with errors (Fig.\ \ref{fig:spec_cos}). 
The FUV lines for which we measured the fluxes are listed in Table \ref{tab:linefluxes}. To this aim, we adopted a Moffat's line profile function that can adequately describe both the core and the wings of the COS lines.
Using both FUV fluxes from COS ion lines and X-ray fluxes measured in RGS spectra, we reconstructed the EMD(T) as described in Sect. \ref{sec:EMD}.

\begin{figure}
    \centering
    \resizebox{\columnwidth}{!}{
     \includegraphics[trim= 1.5cm 13.2cm 4.0cm 4.0cm, clip]{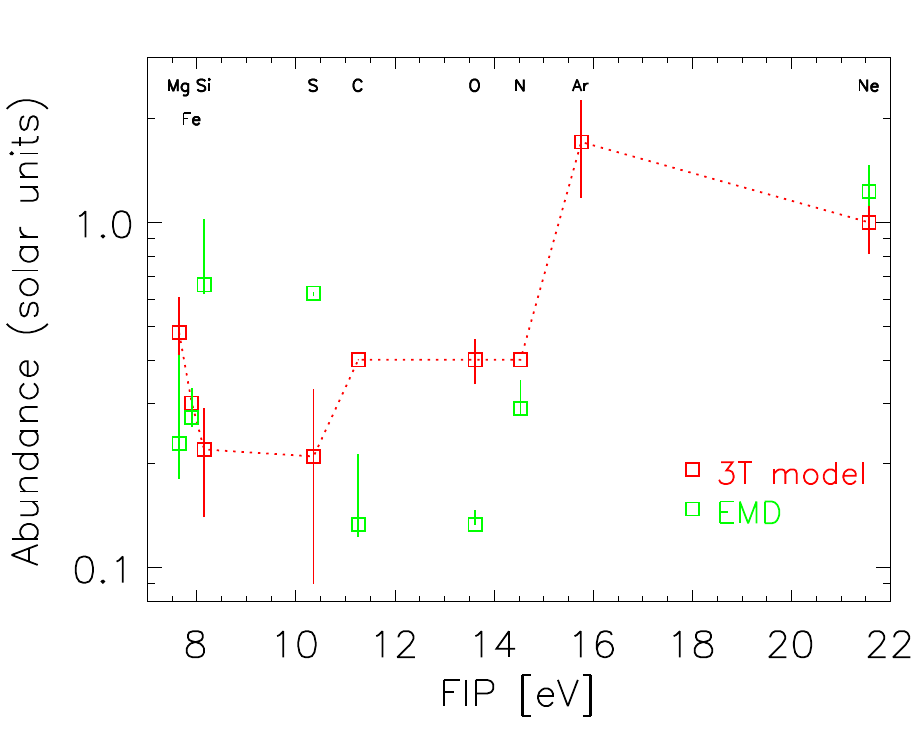}
    }
    \caption{Element abundances versus\ first ionization potential from the global fitting of the EPIC+RGS spectra with a 3T model, and from the emission measure analysis of the COS and RGS line fluxes.  }
    \label{fig:abundances}
\end{figure}

\section{Results 
\label{results}}
\subsection{Global spectral analysis}
\label{sec:globalspec}
HIP\,67522 is an active star with a quite hot corona. The global pn and MOS spectra (Fig.\ \ref{fig:globalspec}) clearly show the line complexes due to H-like and He-like ions of S and Ca, and also the Fe\,XXV He-like line, indicating plasma components with temperatures above 10\,MK.
In effect, the best-fit model includes plasma in the range 5--25\,MK, with the hottest component dominant in terms of volume emission measure (Table \ref{tab:xspecbestfit}).

\begin{table*}[t]
    \caption{Best-fit parameters obtained with {\sc xspec} of spectra during each time interval.}
    \label{tab:xspecbestfit}
    \begin{center}
    \resizebox{0.8\textwidth}{!}{
    \begin{tabular}{lrrrr}\hline\hline
& \multicolumn{1}{c}{Global$^{a}$} & \multicolumn{3}{c}{Quiescent phases$^{b}$} \\
& \multicolumn{1}{c}{\hrulefill} & \multicolumn{3}{c}{\hrulefill} \\
Interval                   &                    & 1                  &  4       & 7                 \\
n$_H$ ($10^{20}$ cm$^{-2}$) &  0.23 ($< 0.80$)   & 0.23 -            &  0.23 -- & 0.23 (1.0--2.0)   \\
T$_1$ ($10^6$\,K)           &  5.3 ( 5.0--5.9)   &  6.5 --           &  6.5 --  &  6.5 (5.5--7.9)   \\
T$_2$ ($10^6$\,K)           & 10.9 (10.6--11.3)   & 11.4 --           & 11.4 --  & 11.4 (10.6--13.6) \\
T$_3$ ($10^6$\,K)           & 25.6 (24.4--26.9)   & 23.0 --           & 23.0 --  & 23.0 (21.1--31.7) \\
Z/Z$_\odot$$^{c}$             &  0.30 (0.26-0.33)  &  0.22 --           &  0.22 -- &  0.22 (0.19--0.26)\\
$\log$ EM$_{1}$            & 52.63 (52.56--52.70) & 52.81 (52.72--52.87) & 52.89 (52.81--52.95) & 52.77 (52.62--53.00)  \\
$\log$ EM$_{2}$            & 52.97 (52.92--53.04) & 53.03 (52.95--53.09) & 52.87 (52.74--52.97) & 53.09 (53.02--53.19)  \\
$\log$ EM$_{3}$            & 53.22 (53.20--53.25) & 52.94 (52.87--53.01) & 53.16 (53.11--53.21) & 53.13 (52.88--53.19)  \\
$\log$  L$_{\rm X}$$^{d}$ & 30.58 (30.57--30.59) & 30.46 (30.45--30.47) & 30.52 (30.51--30.53) & 30.55 (30.54--30.56) \\
$\chi^2$                & 1438                & 48.4                 & 67.8    & 98.2  \\
 D.o.F.                 & 1131                & 53                   & 51      &103    \\
\hline \\
& \multicolumn{2}{c}{First flare$^{e}$} & \multicolumn{2}{c}{Second flare$^{e}$}\\ 
& \multicolumn{2}{c}{\hrulefill} & \multicolumn{2}{c}{\hrulefill}\\
Interval                     & 2                     & 3                   &     5                 &  6  \\
T$_4$ ($10^6$\,K)                  & 111 (86--162)        &  39 (32--47)        & 63 (44--98)          & 53 (32--99)  \\
$\log$  EM$_{4}$              & 53.35 (53.32--53.37) & 53.12 (53.09--53.15)& 53.04 (52.99--53.10) & 52.58 (52.49--52.67)  \\
$\log$  L$_{X,Tot}$ (erg/s)   & 30.87 (30.86--30.88) & 30.71 (30.70--30.72) & 30.75 (30.74--30.76) & 30.64 (30.63--30.65)  \\
L$_{\rm X,flare}$ (erg/s) $^{f}$  & \multicolumn{2}{c}{$6.9 \times 10^{30}$}  & \multicolumn{2}{c}{$3.4 \times 10^{30}$} \\
$\chi^2$                      & 74.7                 &103.1                & 74.5                  & 99.6  \\
 D.o.F.                       & 68                   & 65                  & 52                    & 65  \\
\hline 
\end{tabular} 
}
\end{center}
\tablefoot{
\tablefoottext{a}{Best fit of all EPIC and RGS spectra obtained with 3T VAPEC components.}
\tablefoottext{b}{Best fits of the summed MOS+pn spectra in the quiescent phases were obtained with 3T VAPEC models, where $n_{\rm H}$ and element abundance ratios with respect to iron were fixed at the values of the "Global" best fit.
For intervals 1 and 4, we also fixed the temperatures and the iron abundance found in phase 7, and varied only the emission measures of the three thermal components.}
\tablefoottext{c}{Best-fit value of the iron abundance, linked to the abundances of Al, Ar, and Ni. Abundances of other elements, treated as free parameters, are reported in Sect.\ \ref{results}.}
\tablefoottext{d}{Unabsorbed total luminosity for the quiescent phases are given in the band $0.3-10.0$ keV.}
\tablefoottext{e}{For the flares in phases 2--3 and 5--6 we used a 4T model with three components fixed at the best-fit values found in the pre-flare phases 1 and 4, respectively.}
\tablefoottext{f}{The total flare luminosity in phases 2--3 and 5--6 computed as the total luminosity minus the luminosity of the pre-flare phases 1 and 4, respectively.}
}
\end{table*}

We left the chemical abundances of O, Ne, Mg, Si, S, Ca, and Fe  free to vary (Table \ref{tab:abund}). The abundances of C and N were linked to that of oxygen, while Al, Ar, and Ni remained linked to iron.
The pattern of element abundances versus\ first ionization potential (FIP; Fig.\ \ref{fig:abundances}), suggests a FIP bias typical of young active stars \citep{Maggio+2007,Scelsi+2007}.

\begin{table}[!t]
    \caption{Element abundances derived from the analysis. \label{tab:abund} }
    \centering
    \begin{tabular}{c c | c | c}
    \hline\hline

  & & \multicolumn{1}{c|}{3T global fit} & \multicolumn{1}{c}{EMD analysis} \\
Element & Z & $Ab(Z)/Ab(Z)_\odot$ & $Ab(Z)/Ab(Z)_\odot$ \\
\hline     
C &     6 & \multicolumn{1}{c|}{= O} & 0.13 (0.12$-$0.22) \\
N &     7 & \multicolumn{1}{c|}{= O} & 0.29 (0.28$-$0.35) \\
O &     8 & 0.40 (0.34$-$0.46)      & 0.13 (0.13$-$0.15) \\ 
Ne &   10 & 1.00 (0.81$-$1.20)      & 1.23 (1.11$-$1.47) \\ 
Mg &   12 & 0.48 (0.36$-$0.61)      & 0.23 (0.18$-$0.41) \\ 
Si &   14 & 0.22 (0.14$-$0.29)      & 0.66 (0.62$-$1.02) \\ 
S  &   16 & 0.21 (0.09$-$0.33)      & 0.62 (0.61$-$0.63) \\
Ar &   18 & 1.71 (1.18$-$2.26)    \\ 
Fe &   26 & 0.30 (0.26$-$0.33)      & 0.27 (0.26$-$0.33) \\ 
\hline
    \end{tabular}
\end{table}

The best-fit model also provides a measure of $N_{\rm H} = 2.3 \times 10^{19}$\,cm$^{-2}$, with a $1\sigma$ upper limit of $< 7.9 \times 10^{19}$\,cm$^{-2}$. 
This value agrees with the absorption expected if we assume a mean interstellar hydrogen density of 0.07--0.1\,cm$^{-3}$, typical of the solar neighborhoods. 
In addition, the value is in good agreement with the empirical relationship of \cite{Redfield2000}, which yields $N_{\rm H} \simeq 3.9\times10^{19}$\,cm$^{-2}$.
The $E(B-V) < 0.05$ color excess \citep{Rizzuto2020} provides a looser constraint of $N_{\rm H} < 2 \times 10^{20}$\,cm$^{-2}$.

\subsection{Time-resolved analysis and flares}
\label{sec:flares}
The EPIC light curves show two flares with peaks occurring at approximately 9\,ks and 18\,ks after the start of the exposures. 
The first flare has a relatively symmetrical profile, with the rise and decay phases each having a duration of about 3\,ks, while the second flare shows a more usual profile, with a rapid rise ($t\sim1$\,ks) followed by a slower decay of $t\sim5$\,ks. 

While the flares are clearly visible in soft X-rays, only the 
first one is barely visible in the OM band pass (200--300\,nm) at a planetary phase of $\sim 0.902$, while the second one is much less apparent as its peak occurred during a gap between two subsequent OM exposures. 
For the first flare, the delay between the peak in NUV and in X-rays is about $2.8\pm0.3$\,ks, and is a signature of the Neupert effect \citep{Neupert1968,Hudson1991}. 
The plasma within a loop is first heated and the UV peak probes the rise in temperature; the gas then evaporates through the feet of the loop and fills it, with an increase in the plasma density and the X-ray emission mirroring the rise and peak of the emission measure within the loop
\citep{Namekata2017}. 

In order to perform time-resolved spectroscopy of the quiescent and flaring phases, we divided the light curve into seven intervals, as shown in Fig. \ref{fig:lc} (lower panel). 
For each flare, we considered the rise up to the peak as a first interval where, presumably, we can measure the peak of the plasma temperature (intervals 2 and 5),
and a second interval with the decay after the peak (intervals 3 and 6). 
The quiescent phases were those preceeding the two flares (intervals 1 and 4), and the final phase after the second flare (interval 7). 
The latter has the longest duration, and the spectrum of the combined pn and MOS instruments accumulated during this interval has 43930 counts. 
We fit this spectrum with a model composed of three VAPEC components (see Table \ref{tab:xspecbestfit}), fixing all the abundance ratios with respect to iron at the best-fit values found in the global fit, and leaving only the iron abundance free to vary.
We obtained a cool component at around 6.5\,MK, a mild component at around 11\,MK, and a hot component at 23\,MK.

Since the spectral shape is similar in intervals 1, 4, and 7, for the spectra 1 and 4 we used the same 3T VAPEC model derived from the best fit to the spectrum 7, with only the three emission measures left to vary. 
In this manner, we trace the slight differences ($\sim 20$\% in flux) between these quiescent intervals, while maintaining a minimal number of free parameters, suitable for the lower photon counting statistics in these brief pre-flare intervals.

For the flare segments, we separated the phases of rise + peak and the decay (intervals 2 and 3 for the first flare, and 5 and 6 for the second flare), and used the 3T model from the best fits in the pre-flare phases as a quiescent basal emission, to which we added a fourth APEC component to describe the flaring plasma. 
The maximum temperatures in the flares are about 110\,MK (k$T_4 = 10^{+4}_{-2}$\,keV) and 60\,MK (k$T_4 = 5.4^{+3.0}_{-1.6}$\,keV) during the first and second flare, respectively. 
The presence of very hot plasma is clearly seen as an enhancement of the highly ionized Fe XXV lines at 6.7 keV.

In the hypothesis of a quasi-static cooling of the loop after the flare, it is possible to apply simple diagnostics to infer the semi-length of the loop   \citep{Serio1991, Reale2007, Reale2014}. 
The relevant equations are:
\begin{equation}
L_9=\frac{\tau_{LC} \sqrt{T_7}}{120 F(\zeta)}  ~~~~~~~~ 
\zeta_{min} < \zeta \leq 
\zeta_{max}
\label{eq:lreale}
,\end{equation}
where $L_9$ is the loop semi-length in units of $10^9$ cm, 
$\zeta$ is the slope of the decay in the $1/2 \log(\mathrm{EM})-log(\mathrm{T})$ plane, 
$\tau_{LC}$ is the e-folding time of the light-curve decay, $T_7$ is the maximum
temperature of the flare in units of $10^7$ K and calibrated from the observed maximum temperature $T_{obs}$ inferred from the model best fit to the spectra:
\begin{equation}
T_7 = \xi ~ \frac{T_{obs}^{\eta}}{10^7}
\label{eq:tmax}
.\end{equation}

The correction to the formula of \citet{Serio1991} for a quasi-static decaying loop
in case of residual heating during the decay is 
\begin{equation}
F(\zeta) = \frac{c_a}{\zeta-\zeta_a} + q_a 
,\end{equation}
with parameters estimated from extensive simulations of observations with the EPIC instrumentation
\citep[see][]{Reale2007}.

Using the decay e-folding times (1.5 and 2.2 ks for the first and second flare, respectively) and the maximum observed temperatures (see above), we inferred a semi-length of about 6-$7\times10^{10}$\,cm, similar for both flares.  This value corresponds to about 60--70\% of the stellar radius. Assuming an aspect ratio of 0.1 between the radius and length of the loop, we derived a volume of about 1.8--$2.1\times10^{31}$\,cm$^{3}$. 
From the volume and emission measure $EM_4$ relative to segments 2 and 4, we inferred an electron density of 7--$11\times10^{10}$\,cm$^{-3}$, which appears high when compared to the densities 
recorded during solar flares, but similar to the values of other flares in young stars \citep{Getman2021}.

We estimated the energy released in each flare by summing the net luminosity of the flaring component ($L_{\rm X,flare}$), obtained by subtracting the luminosity in the preflare intervals, multiplied by the duration of each flaring interval (2--3 and 5--6). This resulted in values of $\simeq 2.1\times10^{34}$\,erg and $\simeq 7.6\times10^{33}$\,erg for the two respective flares.  
Similar energies have been detected in other young-planet-hosting stars, such as DS\,Tuc\,A \citep{Pillitteri2022}; these energies are more than one order of magnitude larger than typical flares occurring in the Sun. Flares releasing energies of $E>10^{34}$\,ergs in X-rays 
are often referred to as "super flares" (\citealp[see e.g.,][]{Getman2021}).

The minimum magnetic field required to constrain the plasma in the loop can be calculated as:
\[ B=\sqrt{16\pi k_{\rm B} n_{e,\mathrm{peak}} T_\mathrm{peak}},\]
where $k_{\rm B}$ is the Boltzmann's constant, and $n_{e,\mathrm{peak}}$ and $T_\mathrm{peak}$ are the density and the temperature at the flare peak. 
We inferred minimum fields of $B_{min} \ge460$\,G and $B_{min} \ge270$\,G for the two respective  flares.

\begin{figure}
    \centering
    \resizebox{\columnwidth}{!}{
    \includegraphics[trim=1.8cm 13.5cm 4.0cm 4cm, clip]{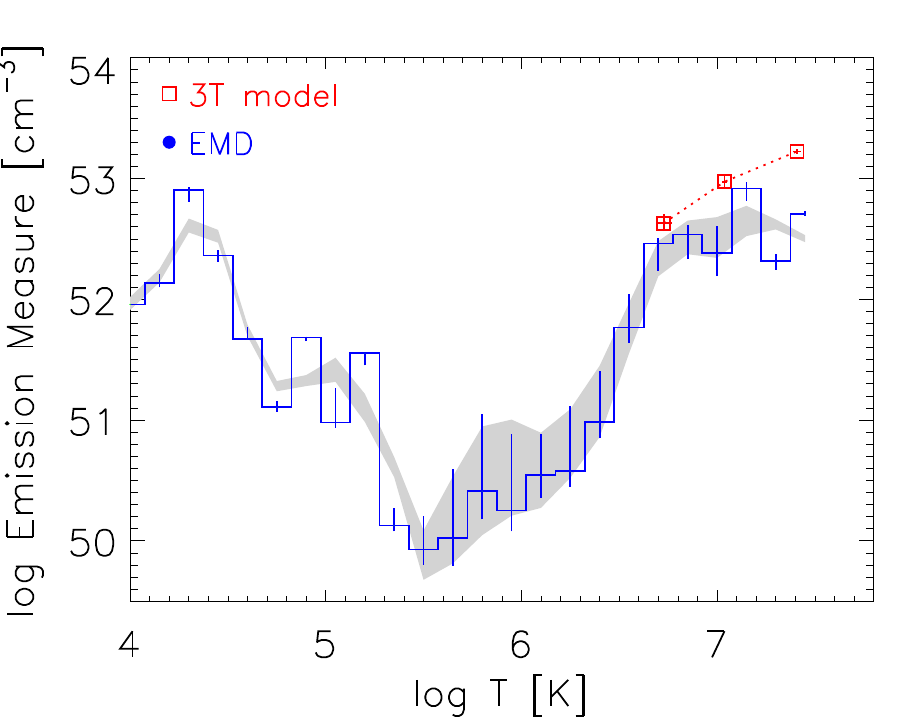}
    }
    \resizebox{\columnwidth}{!}{    \includegraphics{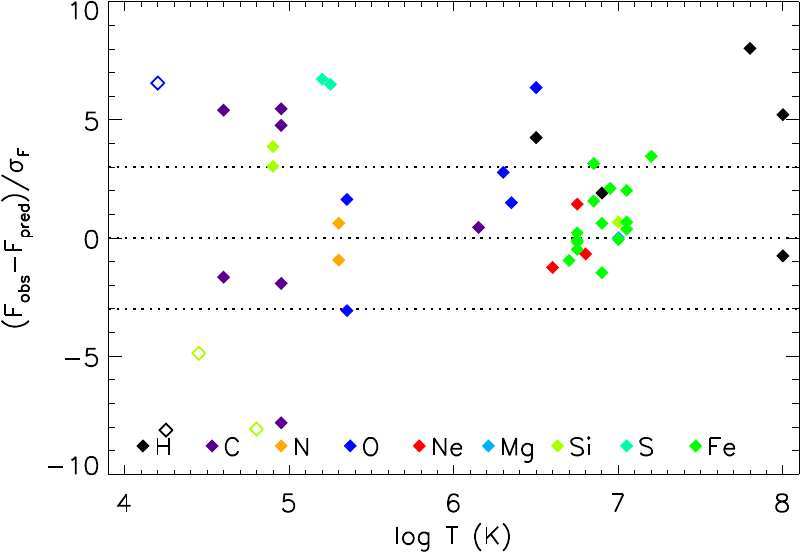}
    }
    \caption{Results of the joint analysis of XMM-Newton/RGS and HST/COS spectra. Top: Plasma EMD vs.\ temperature, compared with the 3-T model (red points) best fitting the global EPIC and RGS spectra.
    The shaded band represents a smoothed $1\sigma$ confidence region.
    Bottom: Differences, in $\sigma$ units, between measured line fluxes and predicted values vs.\ temperature at the peak emissivity. The black "H" symbols represent narrow-band measurements of the X-ray continuum. Empty symbols are lines not used for the EMD reconstruction.
   }
    \label{fig:emd}
\end{figure}

\begin{figure*}
    \centering
    \resizebox{0.495\textwidth}{!}{
\includegraphics{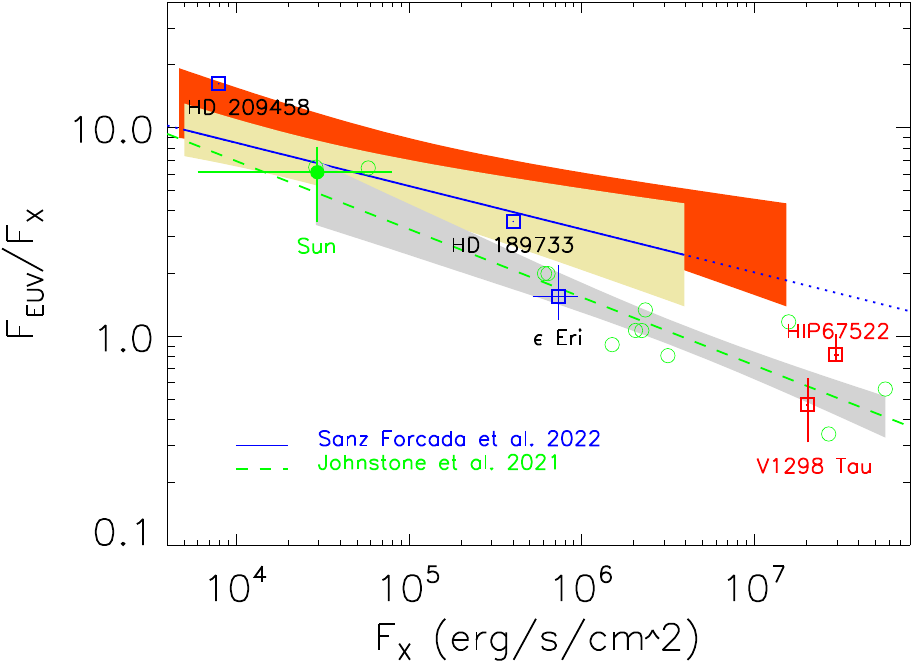}    }
    \resizebox{0.495\textwidth}{!}{
\includegraphics{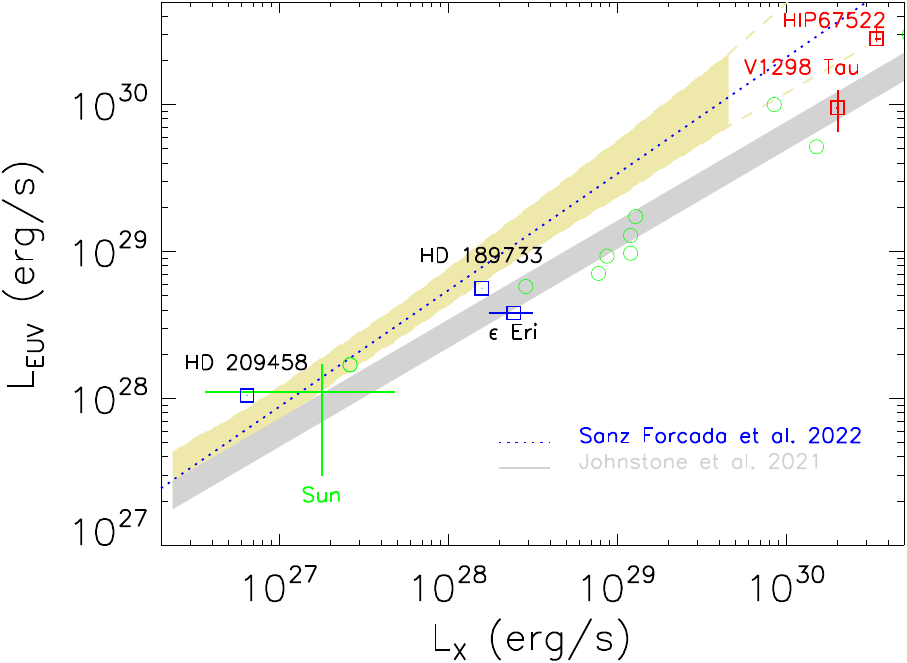}    }
     \caption{X-ray vs.\ EUV scaling laws and benchmark stellar sample. (Left) Measurements of the EUV to X-ray ratio vs.\ X-ray flux at the stellar surface for \hip, the Sun, and other G--K-type stars. The names indicate benchmark stars with exoplanets. Different X-ray (5--100\,\AA) to EUV (100--920\,\AA) scaling laws are shown for comparison: the gray band indicates the 90\% confidence region relative to the J21 scaling law (green dashed line). The golden band is the 90\% confidence region for the SF22 scaling law, assuming stars with radii equal to that of \hip, while the orange band is for stars with 0.7\,R$_\odot$.
     The ranges for the Sun indicate its variability during an entire magnetic cycle.
     (Right) Analogous plot for the scaling laws of the EUV vs.\ X-ray luminosities. The gold dashed lines represent an extrapolation of the SF22 solution, while the gray-shaded area comprises stars with radii in the range 0.7--1.38\,R$_\odot$.
     }
    \label{fig:xuvscaling}
\end{figure*}

\begin{table}
\caption{Emission measure distribution 
}
\label{tab:emd}
\centering
\begin{tabular}{cccc}\\
\hline \hline 
log\,T & log\,EM & log\,EM$_\mathrm{low}$ & log\,EM$_\mathrm{hi}$ \\
(K) & (cm$^{-3}$) & (cm$^{-3}$) & (cm$^{-3}$) \\
\hline
4.00     &      51.96    &      51.88    &      51.98  \\
4.15     &      52.14    &      52.10    &      52.21  \\
4.30     &      52.90    &      52.81    &      52.93  \\
4.45     &      52.36    &      52.31    &      52.41  \\
4.60     &      51.67    &      51.66    &      51.77  \\
4.75     &      51.11    &      51.07    &      51.15  \\
4.90     &      51.68    &      51.65    &      51.69  \\
5.05     &      50.98    &      50.94    &      51.26  \\
5.20     &      51.55    &      51.45    &      51.57  \\
5.35     &      50.13    &      50.08    &      50.27  \\
5.50     &      49.93    &      49.80    &      50.20  \\
5.65     &      50.02    &      49.80    &      50.59  \\
5.80     &      50.41    &      50.18    &      51.05  \\
5.95     &      50.25    &      50.08    &      50.88  \\
6.10     &      50.54    &      50.36    &      50.88  \\
6.25     &      50.58    &      50.45    &      51.11  \\
6.40     &      50.98    &      50.85    &      51.41  \\
6.55     &      51.77    &      51.64    &      52.04  \\
6.70     &      52.46    &      52.24    &      52.51  \\
6.85     &      52.54    &      52.33    &      52.61  \\
7.00     &      52.38    &      52.19    &      52.61  \\
7.15     &      52.92    &      52.82    &      52.97  \\
7.30     &      52.32    &      52.25    &      52.37  \\
7.45     &      52.71    &      52.69    &      52.73  \\
\hline
\end{tabular}
\tablefoot{Logarithm of the volume emission measure (EM) as a function of the plasma temperature. 
Confidence intervals at $1 \sigma$ level are listed in the third and forth columns.}
\end{table}

\subsection{Derivation of the emission measure distribution}
\label{sec:EMD}
The line fluxes from COS and RGS were analyzed with the PoA software in order to derive the emission measure distribution, EMD(T), as a function of plasma temperature.
We adopted Chianti version\ 7.13 as the atomic database for line emissivities. 
For this analysis, we also employed the continuum flux measured in four wavelength regions devoid of lines, derived with XSPEC from the model best fitting the global EPIC+RGS spectrum.
These additional measurements (Table\ \ref{tab:linefluxes}) allow us to constrain the absolute abundances of iron and the other elements.

The plasma EMD(T) versus\ temperature was obtained iteratively with the PoA Monte Carlo Markov Chain procedure by simultaneously varying the emission measure in each temperature bin and the abundances of each element so as to match the measured fluxes. 
To this aim, we considered a subset of measured line fluxes, as reported in Table\ \ref{tab:linefluxes}. In particular, we did not consider density-sensitive lines, lines with uncertain identification, and lines whose fluxes appear mutually incompatible in the hypothesis of collisionally excited plasma \citep{Maggio+2023}.

Considering the temperature ranges covered by the emissivities of the selected lines, we derived the EMD over a temperature grid ranging from $\log T = 4.0$ to $\log T = 7.45$, with $\Delta \log T = 0.15$. 
The results are shown in Fig.\ref{fig:emd} and listed in Table \ref{tab:emd}, together with a plot of the residuals between observed and predicted line fluxes (Table \ref{tab:linefluxes}).

In principle, the use of the average RGS X-ray spectrum acquired during both quiescent and flaring states, in conjunction with the COS FUV lines acquired only during the quiescent state, can introduce a bias in the hot part of the EMD.
Here, we verified that this choice implies a negligible correction of the total flux in the X-ray band (Sect.\ \ref{sec:xuv}).
Moreover, frequent flares are typical of very young and active stars, and should therefore be taken into account, at least in a statistical sense, when evaluating the time-averaged stellar emission level. 
Hence, it is justified to consider the full exposure RGS spectrum, which helps to reduce the uncertainties on the measured X-ray line fluxes.

The element abundances of the model with the highest likelihood and their $1 \sigma$ uncertainties are reported in Table \ref{tab:abund} and shown in Fig.\ \ref{fig:abundances}.
These abundances were found to be compatible with the values from the global fitting of X-ray spectra for the elements N, Ne, Mg, and Fe, while the C, O, S, and Si abundances differ by factors of $\sim 3$. 
Part of the discrepancy can be due to the assumption that the mixing ratios of all the species remain constant in the full range of temperatures explored, while the abundances likely change somewhere between the chromosphere and the corona. 
Moreover, some differences may be due to the two different atomic databases employed for the analysis of the global spectra and of the emission lines.

\begin{figure}
    \centering
    \resizebox{\columnwidth}{!}{
    \includegraphics{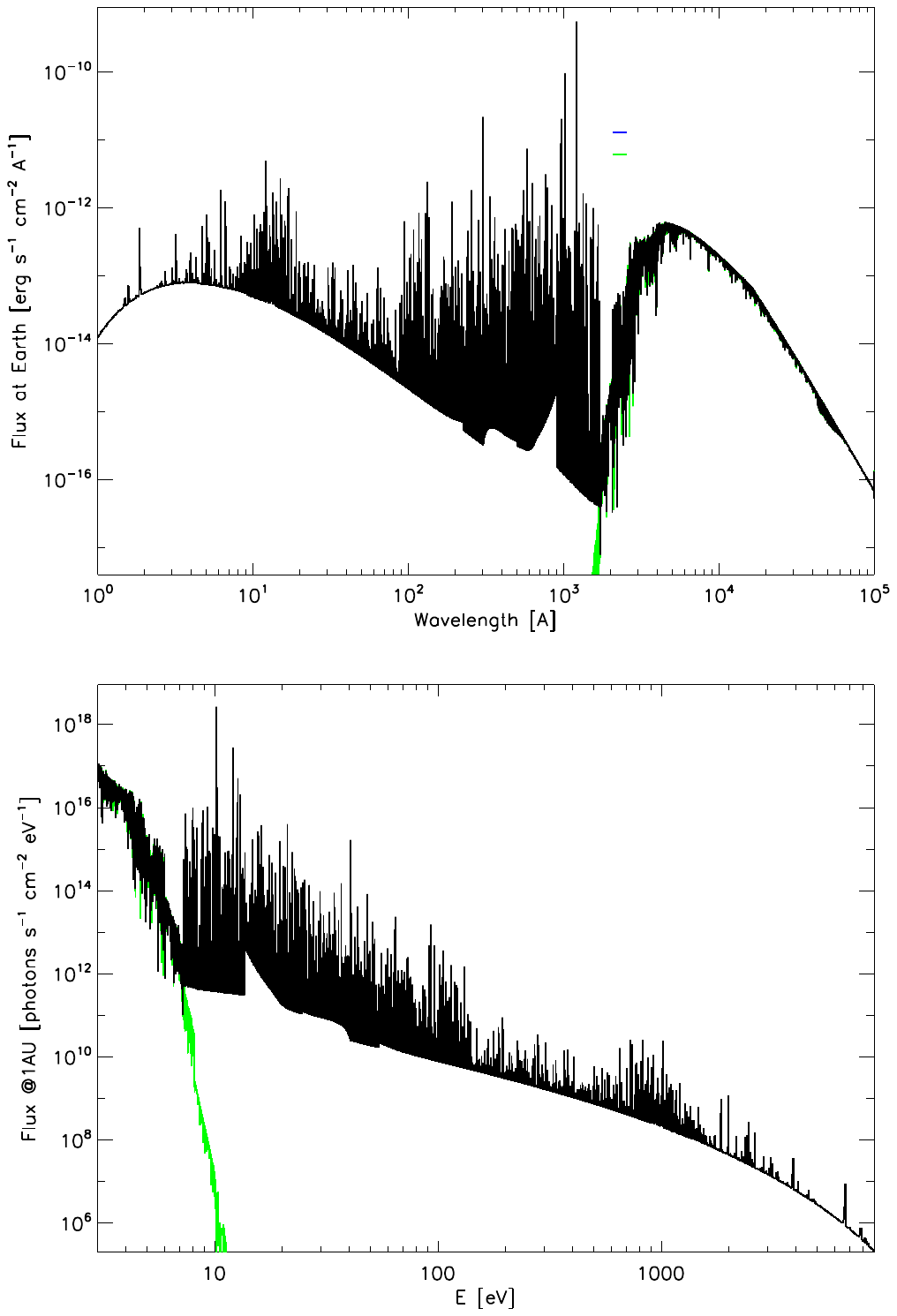}
    }
     \caption{Composite spectrum of \hip\ obtained by joining the Phoenix photospheric spectrum with the XUV spectrum synthesized from the reconstructed plasma emission measure distribution vs. temperature in the chromosphere, transition region, and corona. The upper panel shows the specific flux at Earth, while the bottom panel is the photon flux at a distance of 1\,AU. The Phoenix spectrum resampled\  down to $\sim 1700$\,\AA\  with a wavelength resolution of 1\,\AA \ is shown in green; the XUV spectrum in the range 1--1700\,\AA\ is shown instead with a resolution of 0.01\,\AA. The green and blue segments in the upper panel, at $\sim 200$\,nm, mark the Phoenix model flux and the observed flux integrated over the OM UVM2 band.
}
    \label{fig:fullspec}
\end{figure}

\subsection{Scaling between X-ray and EUV fluxes}
\label{sec:xuv}

We computed synthetic XUV spectra from the EMD(T) presented above (Sect.\ \ref{sec:globalspec}), and then the X-ray luminosity in the 5--100\,\AA\ band, and the EUV luminosity in the 100--920\,\AA\ band. 
We obtained best values of $L_{\rm X} = 3.43 \times 10^{30}$\,erg s$^{-1}$ and $L_{\rm EUV} = 2.80 \times 10^{30}$\,erg s$^{-1}$. 
To evaluate the uncertainties on the luminosities, we determined the 2.5--97.5 percentile ranges  of the luminosity distributions obtained from the Monte Carlo sampling of the EMD(T) and abundances parameter space. 
The ranges are thus
$3.37-3.63 \times 10^{30}$\, erg\,s$^{-1}$ on $L_{\rm X}$ and $2.66-3.56 \times 10^{30}$\, erg\,s$^{-1}$ on $L_{\rm EUV}$. 
The corresponding range of $L_{\rm EUV}/L_{\rm X}$ ratios is 0.76--1.00. 

As we employed the full-exposure RGS spectra, the EMD above $\sim10^6$\,K also takes into account the presence of the flares that occurred during the observation. 
However, their effect on the broad-band X-ray luminosity remains negligible, because the RGS band is softer than the EPIC band and is less sensitive to very hot plasma. 
Indeed, the values of X-ray luminosity derived with EPIC spectra during the quiescent 
intervals 1, 4, and 7 is consistent with the range of X-ray luminosity derived from the synthetic spectrum based on the EMD. 

Figure \ref{fig:xuvscaling} shows our EUV and X-ray measurements for \hip, 
together with a few other benchmark stars with exoplanets already considered in \cite{Maggio+2023}. The sample of G--K dwarfs employed by \citet{Johnstone+2021} (J21) to calibrate their X-ray to EUV scaling law is also shown. 
In the same plot, we repropose this scaling law and the alternative scaling law by \citet[SF22]{SF22}, with the respective 90\% confidence regions. 
In the latter case, where the original relation is based on EUV vs.\ X-ray luminosities, we show two confidence regions in the $F_{\rm EUV}/F_{\rm X}$ vs.\ $F_{\rm X}$ plot, which were obtained assuming two different values for the stellar radius\footnote{For \hip\ we adopted a radius of 1.38\,R$_\odot$ \citep{Rizzuto2020}}. 
We stress that the empirical power laws are based on heterogeneous samples of about 20 single or binary stars, with spectral types from F to M, and observed in X-rays or UV wavelengths at different epochs.

Among the benchmark stars, selected to cover a wide range of activity levels, we included our Sun, with flux ranges derived from \citet[]{Johnstone+2021} and based on observations with the TIMED/SEE mission. 
As intermediate-activity stars, we show the cases of HD\,189733 \citep{Sanz-Forcada2011,Bourrier+2020}) and $\epsilon$\,Eri \citep{Sanz-Forcada2011,Chadney+2015,King+2018}). For the latter, we adopted the X-ray luminosity range derived by \citet{Coffaro+2020}, because the EUV measurement is not simultaneous.
At the high-activity extreme, an interesting comparison case is provided by V\,1298\,Tau, a young planet-hosting star similar to \hip, for which we derived an X-EUV spectrum following the same observation and analysis approach as in the present case \citep{Maggio+2023}. 

\subsection{Planetary irradiation spectrum}
We employed our EMD(T) solution to synthesize the full XUV spectrum in the range 1--1700\,\AA\ (Fig.\ \ref{fig:fullspec}), and extended it to the visible wavelength range by adding the photospheric contribution predicted for a star with $T_{\rm eff} = 5700$\,K, $\log g = 4.0$, and solar metallicity \citep{Rizzuto2020} based on the PHOENIX stellar library \citep{Husser+2013}. 
We also overplot the flux measured with the XMM/OM in the UVM2 band (1830--2790\,\AA), computed on the Vega flux scale: 
taking a range of the OM count rate $\sim 11$--14\,ct\,s$^{-1}$ from Fig.\ \ref{fig:lc}, for \hip\ ($m_{\rm v} = 13.03$) we derived 
a source intensity of $f_{\rm MUV} = 2.39$--$3.05\times 10^{-14}$\,erg\,s$^{-1}$\,cm$^{-2}$\,\AA$^{-1}$, which translates into a flux of 
$F_{\rm MUV} = 2700$--$3500$ \,erg\,s$^{-1}$\,cm$^{-2}$ ($2.7-3.5$ W m$^{-2}$) at the planet distance ($a\sim0.076$ AU).
We make the synthetic spectrum of \hip\ in FITS format  available for download from the Zenodo archive\footnote{URL address {https://doi.org/10.5281/zenodo.13713288}}. 
The file contains the tables with the energy, wavelength, and flux at 1\,AU along with the EMD(T) used to synthesize it and the element abundances $Ab(Z)/Ab(Z)_\odot$.

\subsection{Planetary photoevaporation}
\label{sec:evap}

\begin{table*}
    \scriptsize
    \caption{Results of the photo-evaporation modeling.}
    \label{tab:evap}     
    \centering              
    \begin{tabular}{c r|r c c|l c}       
        \hline    
    \noalign{\smallskip}
    Radius & Mass & Core mass& Core radius & $f_{\rm atm}$ & Mass-loss rate & Timescale \\     
    ($R_\oplus$) & ($M_\oplus$) & ($M_\oplus$) & ($R_\oplus$) & (\%) & (g/s)  & (Myr) \\    
     \noalign{\smallskip} 
     \hline
    \noalign{\smallskip}
    & & \multicolumn{3}{c}{Low opacity} & \multicolumn{2}{c}{ } \\
     10.0 & 57 & 37 & 3.0 & 36 & $2.5 \times 10^{12}$ & 580 \\
     10.0 & 318 & 154 & 3.9 & 52 & $5.4\times 10^{9}$ & $\gg 10$\,Gyr  \\
    \hline   
    \noalign{\smallskip}
    & & \multicolumn{3}{c}{High opacity} & \multicolumn{2}{c}{ } \\
    10.0 & 57  & 47 & 3.1 & 18 & $2.5 \times 10^{12}$  & 280  \\  
     10.0 & 318  & 242 & 4.2 & 24 & $5.4 \times 10^{9}$  & $\gg 10$\,Gyr  \\ 
    \hline   
    \end{tabular}
\end{table*}

We employed the measured X-ray and EUV fluxes to compute the atmospheric mass-loss rate of the planet HIP\,67522b using the analytical approximation based on the ATES hydrodynamic code \citep{Caldiroli+2021,Caldiroli+2022}.
In order to evaluate possible evaporation timescales, we also need an educated guess of the atmospheric mass fraction, which we derived from the planetary core--envelope models by \cite{Fortney2007} and \cite{LopFor14}. 
To this aim, we assumed an ice/rock composition of the core of 25\%/75\%.
In practice, most of the uncertainty in modeling the planet structure rests in the planetary mass, which is poorly constrained (0.18--4.6 $M_\mathrm{J}$ \citealt{Rizzuto2020}). Considering that the planetary radius is just 10\% lower than the size of Jupiter, we explored four cases corresponding to two values of mass, 
$M_{\rm p} = 0.18$\,M$_\mathrm{J}$ or 1\,$M_\mathrm{J}$, and two possible values of atmospheric metallicity, yielding solar or enhanced opacities (see \citealt{LopFor14}).

At the nominal radius of HIP\,67522b, the system of equations that describe the planetary structure allows only one solution with a relatively large and massive core. 
Table \ref{tab:evap} shows the values of the core mass, core radius, and atmospheric mass fraction at present age in the four cases. 
Assuming that the core mass and size remain constant in time, the envelope radius and atmospheric mass fraction evolve in response to gravitational contraction and to photoevaporation, the latter driven by the X-EUV irradiation. 

At its present age, the X-EUV flux (5-920\,\AA) is found to be $\sim 3.9 \times 10^5$\,erg\,s$^{-1}$\,cm$^{-2}$ at the planet, which is relatively high, and implies a strong energy loss due to advective and radiative cooling if the gravitational potential is low enough \citep{Caldiroli+2022}. This is our low-mass case for HIP\,67522b, that is,\ in a low-gravity regime of the atmospheric hydrodynamic outflow, which occurs when the volume-averaged mean excess energy due to photo-heating exceeds the gravitational binding energy. As a consequence, the ATES model predicts a photoevaporation efficiency of $\eta \sim 14$\% with respect to the energy-limited threshold \citep{Erkaev+2007}. 
The efficiency is even lower in the high-mass case, that is,\ in a high-gravity regime: $\eta \sim 0.2$\%. This is because advective cooling and Ly$\alpha$ energy losses dominate over adiabatic expansion and cooling. 
This difference leads to a mass-loss rate of $\sim 10^{-2}$\,\mearth/Myr in the low-mass case, or $\sim 3 \times 10^{-5}$\,\mearth/Myr in the high-mass case (Table\, \ref{tab:evap}). 
Depending on the planetary mass and structure, the instantaneous e-folding evaporation timescale will be relatively short (300 -- 600\,Myr) in the low-mass case, or extremely long ($\gg 10$\,Gyr) in the high-mass case.

The actual long-term evolution of the planetary mass and radius depends on the rate of decrease in the stellar activity and its X-EUV emission with age, and on the relative role of mass loss with respect to gravitational contraction during the evolution (see e.g.,\ \citealt{Mantovan+2024,Damasso+2024}).
Such a highly refined modeling is premature for HIP\,67522b, given the poor knowledge of the actual planetary mass.

\section{Discussion and Conclusions}\label{conclusions}

In this paper, we present the results from simultaneous observations of \hip\ with \xmm\ and HST in X-ray and FUV bands.
The quiescent X-ray luminosity of this target, $L_{\rm x} \sim 3 \times 10^{30}$\,erg\,s$^{-1}$, is very high and in line with that expected for a $1.2$\,M$_\odot$ star at an age of 15--18\,Myr \citep{Johnstone+2021}.
The high activity level is also mirrored by the hardness of the X-ray spectrum, which is due to the presence of plasma at temperatures exceeding 20\,MK.

The star also exhibited two moderately bright flares separated by $\sim 9$\,ks; these flares released energies of between $8 \times 10^{33}$ and $2 \times 10^{34}$\,erg.
In this respect, they appear similar in energy and timing to a couple of flares detected in DS\,Tuc\,A, which is 40\,Myr old and of similar stellar mass \citep{Pillitteri2022}.
We speculate that such twin flares in young coronae could be triggered in the same loop or in adjacent loops due to the packed structuring of the magnetic field. 
At the distance of the planet, the X-ray flux received during the two flares was 
$\sim 790$ and $\sim 630$\,W\,m$^{-2}$. 

From a different perspective, \cite{Ilin+2024} recently presented a survey of optical flares in planet-hosting stars detected during the \textit{Kepler} and TESS space missions. In particular, they searched for flaring events clustered in orbital phase, as a proxy of star--planet magnetic interactions. \hip\ was found to be the target most likely characterized by this kind of behavior based on the distribution of 12 flares observed in three sectors of the TESS sky coverage.

\begin{figure}
    \centering
    \resizebox{0.48\textwidth}{!}{\includegraphics{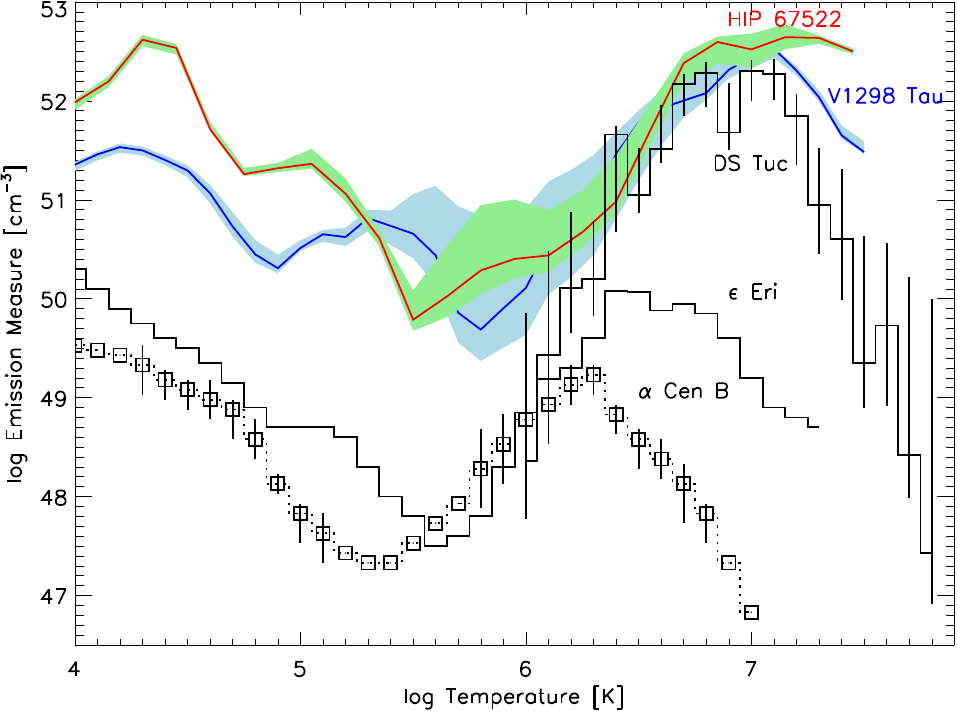}}
    \caption{Comparison of plasma emission measure distribution vs.\ temperature for \hip\ and four other G--K stars with different activity levels (see text). A polynomial smoothing (order 2, 0.6 dex in width) was applied to the low and high $1 \sigma$ boundaries of the EMD solutions for \hip\ (Sect.\ \ref{sec:EMD} and for V\,1298\,Tau \citep{Maggio+2023}.}
    \label{fig:dems}
\end{figure}

We derived the volume emission measure distribution of the plasma vs.\ temperature, which provides\ an overall description of the average thermal structure of the plasma from the upper chromosphere to the corona.
In Fig.\ \ref{fig:dems} we compare the EMD of \hip\ to those of a few other G-K stars with different activity levels and already considered in \cite{Maggio+2023}. In particular, we note the similarity between the EMDs of \hip, V\,1298\,Tau, and DS\,Tuc in the corona ($T > 10^6$\,K), along with an excess of emission measure for \hip\ in the chromosphere and transition region.

The EUV/X-ray flux ratio for \hip, inferred from the synthetic spectrum based on the average EMD, 
places this young active star in between the two alternative scaling laws of \cite{Johnstone+2021} and \cite{SF22}.
This $F_{\rm EUV}/F_{\rm X}$ ratio is larger, at the $1 \sigma$ level, than the value we derived for V1298\,Tau, which was also based on simultaneous X-ray and FUV observations analyzed with a methodology similar to the present case \citep{Maggio+2023}. 
This result is explained by the different shapes of the EMDs, and it suggests that there is an intrinsic uncertainty, by at least a factor 2, on the EUV flux derived for other stars, when only X-ray measurements are available, depending on the assumed scaling law.
An uncertainty of similar size can be systematic in origin, and due to different possible methodologies employed to reconstruct the EMD(T) and hence to estimate indirectly the EUV flux from available X-ray and FUV spectra \citep{Maggio+2023}. 

We employed the total X-EUV flux to study the photoevaporation of \hip~b using the ATES code \citep{Caldiroli+2022}.
However, the large uncertainty in the planetary mass (0.18--4.6 M$_{\rm J}$) yields a relatively loose range of possible mass-loss rates.
For a mass of $> 1 M_{\rm J}$, we estimate a rate of $\log \dot M < 9.7$ (g/s), which implies an evaporation e-folding timescale of $\gg 10$\,Gyr. 
For the lower limit on mass of $0.2 M_{\rm J}$, we determine an instantaneous mass-loss rate of $\log \dot M \sim 12.4$ (g/s),
which amounts to $\simeq 4 \times 10^{-5} M_{\rm J}$/Myr.
Considering an atmospheric mass fraction in the range of 18--36\%, the low-mass case yields an e-folding evaporation timescale within the age of stars in the Hyades open cluster.
However, the actual planetary evolution is highly nonlinear, because the planetary high-energy irradiation should decrease by about a factor 5 within 600\,Myr and by a factor 10 within 2\,Gyr due to the natural decay of stellar activity \citep{Johnstone+2021}. On the other hand, for a planet in the low-gravity regime (Sect.\ \ref{sec:evap}), a lower X-EUV flux leads to higher photoevaporation efficiency.
We defer more detailed simulations of the photoevaporation history of \hip\ --- as already performed for V\,1298\,Tau \citep{Maggio+2022}--- to a time when the planet mass has been assessed with future radial velocity follow-up campaigns. 
As a final result, we derived a synthetic XUV spectrum for \hip,\, which will be useful for future studies of photochemistry and photoevaporation in the atmosphere of the Jupiter-size planet hosted by this young solar-type analog.

\bibliographystyle{aa}

\begin{acknowledgements}
The authors acknowledge partial support by the project Exo-planetary Cloudy Atmospheres and Stellar High energy (Exo-CASH) funded by MUR - PRIN 2022 (grant no. 2022J7ZFRA), and the ASI-INAF agreement 2021-5-HH.0.
A.M. also acknowledges support by the project HOT-ATMOS (PRIN INAF 2019).
D.L. acknowledges contributions from Bando di Ricerca Fondamentale INAF—MINI-GRANTS di RSN 2 and STILES (Strengthening the Italian leadership in ELT and SKA).
This work is based on observations obtained with XMM-Newton, an ESA science mission with instruments and contributions 
directly funded by ESA Member States and NASA, and on observations made with the NASA/ESA Hubble Space Telescope, 
obtained from the Space Telescope Science Institute, which is operated by the Association of 
Universities for Research in Astronomy, Inc., under NASA contract NAS 5–26555. These observations are associated with HST program 16901.
\end{acknowledgements}

\appendix
\section{Data table}
\begin{table*}
\caption{Measured X-ray and FUV line fluxes of HIP~67522.}
\label{tab:linefluxes}
\begin{center}
\begin{tabular}{@{\hspace{-1mm}}r@{$\div$}llcr@{$\;\pm\;$}lcc}
\hline\hline
\multicolumn{2}{c}{$\lambda^{a}$} & Ion & $\log T_{\rm max}$$^{b}$ & \multicolumn{2}{c}{Flux$^{c}$} & $EMD$$^{d}$ & $((F_{obs}-F_{pred})/\sigma)$$^{e}$ \\
\hline
\multicolumn{ 8}{c}{Line fluxes} \\
\hline
\multicolumn{2}{r}{    6.65\hspace{5mm} } &                                                                 \ion{Si}{XIII}\, &  7.00 &       191.4 &       118.7 & $\ast$ &        0.67 \\
\multicolumn{2}{r}{    8.42\hspace{5mm} } &                                                   \ion{Mg}{XII}\,\ion{Mg}{XII}\, &  7.00 &        47.8 &        50.3 & $\ast$ &        0.02 \\
\multicolumn{2}{r}{   10.24\hspace{5mm} } &                                                       \ion{Ne}{X}\,\ion{Ne}{X}\, &  6.80 &        50.9 &        32.8 & $\ast$ &       -0.67 \\
\multicolumn{2}{r}{   11.74\hspace{5mm} } &                                                                \ion{Fe}{XXIII}\, &  7.20 &       253.5 &        51.6 & $\ast$ &        3.46 \\
\multicolumn{2}{r}{   12.13\hspace{5mm} } &                                       \ion{Ne}{X}\,\ion{Ne}{X}\,\ion{Fe}{XVII}\, &  6.75 &       601.7 &        67.4 & $\ast$ &        1.44 \\
\multicolumn{2}{r}{   12.28\hspace{5mm} } &                                                  \ion{Fe}{XXI}\,\ion{Fe}{XVII}\, &  7.05 &       149.1 &        50.6 & $\ast$ &        0.38 \\
\multicolumn{2}{r}{   12.83\hspace{5mm} } &                        \ion{Fe}{XX}\,\ion{Fe}{XX}\,\ion{Fe}{XX}\,\ion{Fe}{XXI}\, &  7.05 &       182.6 &        46.5 & $\ast$ &        2.01 \\
\multicolumn{2}{r}{   13.45\hspace{5mm} } &                                     \ion{Ne}{IX}\,\ion{Fe}{XIX}\,\ion{Fe}{XIX}\, &  6.60 &       120.9 &        46.0 & $\ast$ &       -1.24 \\
\multicolumn{2}{r}{   13.52\hspace{5mm} } &                      \ion{Ne}{IX}\,\ion{Fe}{XIX}\,\ion{Fe}{XIX}\,\ion{Fe}{XXI}\, &  7.00 &       104.9 &        44.9 & $\ast$ &       -0.07 \\
\multicolumn{2}{r}{   13.82\hspace{5mm} } &                                                  \ion{Fe}{XVII}\,\ion{Fe}{XIX}\, &  6.90 &        59.2 &        28.7 & $\ast$ &        0.62 \\
\multicolumn{2}{r}{   14.20\hspace{5mm} } &                                               \ion{Fe}{XVIII}\,\ion{Fe}{XVIII}\, &  6.90 &        74.7 &        29.7 & $\ast$ &       -1.47 \\
\multicolumn{2}{r}{   15.01\hspace{5mm} } &                                                                 \ion{Fe}{XVII}\, &  6.75 &       259.1 &        32.3 & $\ast$ &       -0.09 \\
\multicolumn{2}{r}{   15.21\hspace{5mm} } &                                    \ion{Fe}{XIX}\,\ion{O}{VIII}\,\ion{O}{VIII}\, &  6.95 &        87.2 &        33.5 & $\ast$ &        2.10 \\
\multicolumn{2}{r}{   15.26\hspace{5mm} } &                                                                 \ion{Fe}{XVII}\, &  6.75 &        80.6 &        31.6 & $\ast$ &        0.22 \\
\multicolumn{2}{r}{   16.00\hspace{5mm} } &                                  \ion{Fe}{XVIII}\,\ion{O}{VIII}\,\ion{O}{VIII}\, &  6.85 &        96.9 &        26.6 & $\ast$ &        1.56 \\
\multicolumn{2}{r}{   16.07\hspace{5mm} } &                                                                \ion{Fe}{XVIII}\, &  6.85 &       136.6 &        27.5 & $\ast$ &        3.16 \\
\multicolumn{2}{r}{   16.78\hspace{5mm} } &                                                                 \ion{Fe}{XVII}\, &  6.75 &       131.7 &        26.1 & $\ast$ &       -0.48 \\
\multicolumn{2}{r}{   17.05\hspace{5mm} } &                                                                 \ion{Fe}{XVII}\, &  6.75 &       177.0 &        33.3 & $\ast$ &       -0.17 \\
\multicolumn{2}{r}{   17.10\hspace{5mm} } &                                                                 \ion{Fe}{XVII}\, &  6.70 &       108.0 &        39.7 & $\ast$ &       -0.95 \\
\multicolumn{2}{r}{   18.63\hspace{5mm} } &                                                                   \ion{O}{VII}\, &  6.35 &        33.2 &        21.2 & $\ast$ &        1.50 \\
\multicolumn{2}{r}{   18.97\hspace{5mm} } &                                                   \ion{O}{VIII}\,\ion{O}{VIII}\, &  6.50 &       379.4 &        40.1 & $\ast$ &        6.37 \\
\multicolumn{2}{r}{   21.60\hspace{5mm} } &                                                                   \ion{O}{VII}\, &  6.30 &       116.3 &        38.2 & $\ast$ &        2.78 \\
\multicolumn{2}{r}{   33.73\hspace{5mm} } &                                                       \ion{C}{VI}\,\ion{C}{VI}\, &  6.15 &        23.8 &        28.6 & $\ast$ &        0.45 \\
\multicolumn{2}{r}{ 1174.93\hspace{5mm} } &                                                                   \ion{C}{III}\, &  4.95 &        15.5 &         1.9 & $\ast$ &       -1.92 \\
\multicolumn{2}{r}{ 1175.26\hspace{5mm} } &                                                                   \ion{C}{III}\, &  4.95 &        23.4 &         2.5 & $\ast$ &        5.47 \\
\multicolumn{2}{r}{ 1175.71\hspace{5mm} } &                                       \ion{C}{III}\,\ion{C}{III}\,\ion{C}{III}\, &  4.95 &        64.8 &         2.5 & $\ast$ &       -7.82 \\
\multicolumn{2}{r}{ 1176.37\hspace{5mm} } &                                                                   \ion{C}{III}\, &  4.95 &        23.5 &         1.8 & $\ast$ &        4.77 \\
\multicolumn{2}{r}{ 1199.14\hspace{5mm} } &                                                                     \ion{S}{V}\, &  5.20 &        10.8 &         1.1 & $\ast$ &        6.72 \\
\multicolumn{2}{r}{ 1200.15\hspace{5mm} } &                                                                             ...  &  ...  &         7.0 &         1.0 &        &       10.07 \\
\multicolumn{2}{r}{ 1206.50\hspace{5mm} } &                                                                  \ion{Si}{III}\, &  4.80 &       142.6 &         3.6 &        &     -113.63 \\
\multicolumn{2}{r}{ 1218.35\hspace{5mm} } &                                                                     \ion{O}{V}\, &  5.35 &        13.1 &         1.4 & $\ast$ &       -3.07 \\
\multicolumn{2}{r}{ 1238.82\hspace{5mm} } &                                                                     \ion{N}{V}\, &  5.30 &        31.3 &         1.4 & $\ast$ &        0.62 \\
\multicolumn{2}{r}{ 1242.81\hspace{5mm} } &                                                                     \ion{N}{V}\, &  5.30 &        14.7 &         1.1 & $\ast$ &       -0.93 \\
\multicolumn{2}{r}{ 1264.74\hspace{5mm} } &                                                                   \ion{Si}{II}\, &  4.45 &         5.8 &         0.8 &        &      -67.89 \\
\multicolumn{2}{r}{ 1289.34\hspace{5mm} } &                                                                             ...  &  ...  &        29.1 &         6.8 &        &        6.09 \\
\multicolumn{2}{r}{ 1298.95\hspace{5mm} } &                                                   \ion{Si}{III}\,\ion{Si}{III}\, &  4.80 &         6.8 &         1.1 &        &       -8.09 \\
\multicolumn{2}{r}{ 1309.28\hspace{5mm} } &                                                                   \ion{Si}{II}\, &  4.45 &         3.7 &         2.1 &        &       -4.87 \\
\multicolumn{2}{r}{ 1334.53\hspace{5mm} } &                                                                    \ion{C}{II}\, &  4.60 &        83.7 &         3.3 & $\ast$ &        5.41 \\
\multicolumn{2}{r}{ 1335.71\hspace{5mm} } &                                                                    \ion{C}{II}\, &  4.60 &       123.2 &         3.6 & $\ast$ &       -1.66 \\
\multicolumn{2}{r}{ 1351.44\hspace{5mm} } &                                                                             ...  &  ...  &         5.3 &         1.1 & $\ast$ &        6.50 \\
\multicolumn{2}{r}{ 1354.07\hspace{5mm} } &                                                                  \ion{Fe}{XXI}\, &  7.05 &         7.7 &         1.3 & $\ast$ &        0.67 \\
\multicolumn{2}{r}{ 1355.60\hspace{5mm} } &                                                                             ...  &  ...  &         6.2 &         1.2 &        &        6.56 \\
\multicolumn{2}{r}{ 1371.30\hspace{5mm} } &                                                                     \ion{O}{V}\, &  5.35 &         2.5 &         1.0 & $\ast$ &        1.63 \\
\multicolumn{2}{r}{ 1393.76\hspace{5mm} } &                                                                   \ion{Si}{IV}\, &  4.90 &        92.1 &         2.6 & $\ast$ &        3.87 \\
\multicolumn{2}{r}{ 1402.77\hspace{5mm} } &                                                                   \ion{Si}{IV}\, &  4.90 &        48.4 &         2.6 & $\ast$ &        3.03 \\
\hline
\multicolumn{ 8}{c}{Total fluxes in selected wavelength intervals} \\
\hline
            $[   27.55 $ & $    30.24]$ &                                                                                  &  6.50 &       436.0 &         6.8 & $\ast$ &        4.25 \\
            $[    8.49 $ & $     8.79]$ &                                                                                  &  7.80 &       214.0 &         3.0 & $\ast$ &        8.03 \\
            $[    5.17 $ & $   123.98]$ &                                                                                  &  6.90 &     19500.0 &       979.7 & $\ast$ &        1.90 \\
            $[    4.13 $ & $     5.17]$ &                                                                                  &  8.00 &       998.0 &        25.5 & $\ast$ &       -0.75 \\
            $[    2.48 $ & $     4.13]$ &                                                                                  &  8.00 &      1600.0 &        40.0 & $\ast$ &        5.22 \\
\hline
\end{tabular}
\end{center}
\tablefoot{
\tablefoottext{a}{Wavelengths (\AA).}
\tablefoottext{b}{Temperature (K) of maximum emissivity.}
\tablefoottext{c}{Observed fluxes (${\rm 10^{-16}\,erg\,s^{-1}\,cm^{-2}}$) with uncertainties at the 68\% confidence level.}
\tablefoottext{d}{Flux measurements selected for the EMD reconstruction.}
\tablefoottext{e}{Comparison between observed and predicted line fluxes.}
}
\normalsize
\end{table*}

\end{document}